# RASSA: Resistive Pre-Alignment Accelerator for Approximate DNA Long Read Mapping

**Roman Kaplan**
Technion, Israel Institute of Technology

**Leonid Yavits**
Technion, Israel Institute of Technology

**Ran Ginosar**
Technion, Israel Institute of Technology

DNA read mapping is a computationally expensive bioinformatics task, required for genome assembly and consensus polishing. It requires to find the best-fitting location for each DNA read on a long reference sequence. A novel resistive approximate similarity search accelerator, RASSA, exploits charge distribution and parallel in-memory processing to reflect a mismatch count between DNA sequences. RASSA implementation of DNA long read pre-alignment outperforms the state-of-art solution, minimap2, by 16-77× with comparable accuracy and provides two orders of magnitude higher throughput than GateKeeper, a short-read pre-alignment hardware architecture implemented in FPGA.

Constructing human DNA sequence in real time is paramount to development of precision medicine[1] and on-site pathogen detection of disease outbreaks[2]. Single-molecule, real-time sequencing from Pacific Biosciences[3] (PacBio) and Oxford Nanopore Technologies[4] (ONT) are new technologies that can produce long reads within minutes, potentially enabling real time genomic analysis. However, long read DNA sequencing poses new challenges. First, long reads contain many thousands of base pairs (*bps*). Second, long reads tend to exhibit about 15-20% insertion, deletion (*indel*) and substitution errors[3,4].

To construct a complete host sequence, in case a reference sequence exists (from a previously sequenced organism), long reads are mapped to high-similarity locations of the reference sequence. Determining the edit distance between every mapped read and the reference sequence requires a computationally intensive local alignment procedure (e.g., Smith-Waterman[4]). Its computational time complexity is typically $O(nm)$ for two sequences with lengths $n$ and $m$. Reference sequences vary from several millions to billions of bps. It is therefore computationally prohibitive to perform optimal alignment of every long read with the entire reference sequence.

Read mappers (e.g., minimap[6], minimap2[7]) find regions of high similarity (*mappings*) between reads or between a read and a reference sequence, followed by an alignment step to determine the exact edit distance and verify that the mapping is correct. In case that a pre-alignment algorithm identifies a specific region in the reference suitable for mapping, the alignment can be performed only on that region, reducing alignment's duration and resource requirements[8]. Therefore, read mapping can be viewed as a two-step process: (1) pre-alignment filtering and (2) accurate alignment verification. The pre-alignment step reduces the problem size for aligners by narrowing the regions to ones with potentially high-scoring alignment.

Existing pre-alignment hardware solutions[9,10] target short reads (up to several hundred bps) which contain a small number of indel and substitution errors (less than 5%) and have a different error profile than that of PacBio or ONT long reads[3,4]. High edit distance threshold is required for mapping long but error-prone reads. However, current solutions[9] have high false positive rates when the edit distance is high (i.e., greater than 15). Thus, the current solutions for short reads are not applicable for long reads.

Approximate computing techniques are known to trade accuracy for speed or energy efficiency. In case of long reads, multiple errors are a natural part of the sequencing output. Therefore, DNA long read pre-alignment filtering inherently tolerates the imprecision.

With the end of Dennard scaling and the slowdown of Moore's law, novel hardware solutions for data intensive problems are researched. Emerging technologies such as resistive memories enable new architectures with better performance and energy efficiency. Resistive approximate Hamming distance solutions exist[11]. However, these do not provide the parallelism required to support a high throughput application such as DNA read mapping.

In this work, we present RASSA, a Resistive Approximate Similarity Search Accelerator architecture for DNA long read pre-alignment filtering. RASSA is a massively parallel in-memory processor, facilitating simultaneous compare of a long read with a reference sequence. The output of RASSA are locations on the reference sequence, where alignment may result in high score. The key performance breakthrough of RASSA is achieved by applying the similarity search in parallel to the entire reference. While the complexity of alignment is $O(mn)$, RASSA employs in-memory parallel computing on $O(m)$ memory cells to reduce computation time to $O(n)$, where $m$ and $n$, are read and reference lengths, respectively.

RASSA employs resistive elements, memristors, serving at the same time as single bit storage elements and comparators. Additional evaluation transistors translate mismatch scores into voltage levels, which are converted to digital values using Analog to Digital Converters (ADC). Further processing determines the most likely overlap candidates.

This work makes the following contributions:

1. RASSA, an in-memory processing resistive approximate similarity search accelerator, is introduced. The parallel processing architecture is presented bottom-up, from the memristor-based bitcell to base pair encoding and up to a complete RASSA system;

2. RASSA based implementation of long read pre-alignment filtering is developed;

3. Evaluation of RASSA's pre-alignment filtering accuracy and comparative analysis of its execution time and throughput is conducted.

# BACKGROUND

The following two subsections provide concise background on the problem; DNA read mapping, and the memristor device technology.

## DNA Read Mapping

DNA sequencers output fragmented regions of DNA called *reads*. The reads originate from random locations in the genome and may be overlapping. If a DNA sequence from the same species exists, the DNA reads are matched to such existing (reference) sequence (Figure 1a). The main assumption behind this approach is that the reference and the reads originate from the same species, and therefore contain small number of differences (typically less than 1%). Since the main computational effort in this process is placing the reads correctly to the reference sequence, it is called read mapping.

One common read mapping technique is individual mapping of fixed length segments (seeds) of a read[6,7]. Each seed is used as a key in a pre-calculated hash table, which values are the seed locations in the reference sequence. Such locations are then extended and a precise alignment is performed. Once all reads are placed in their positions, the new sequence is constructed from the overlapping regions of the reads. A disadvantage of this approach is its high memory requirements, long running time and complexity.

## Resistive Memories

Resistive memories store information by modulating the resistance of nanoscale storage elements, called *memristors*. They are nonvolatile, free of leakage power, and emerge as potential alternatives to charge-based memories[12], including NAND flash. Memristors are two-terminal devices, where the resistance of the device is changed by the electrical current or voltage. The resistance of the memristor is bounded by a minimum resistance $R_{ON}$ (low resistive state) and a maximum resistance $R_{OFF}$ (high resistive state).

# RASSA: RESISTIVE APPROXIMATE SIMILARITY SEARCH ACCELERATOR

Analyzing long read alignments to a reference sequence reveals long fragments of indel-free high-similarity regions (as in Figure 1a). These regions usually contain tens of bps with few substitutions, and can sometimes reach hundreds of bps (for example, in case of high-accuracy CCS reads[3]). This has motivated us to use simple Hamming distance as a heuristic to find the overlap positions of long reads against a reference sequence. To overcome indels and find high-similarity sections, all possible overlap positions of a read against a reference are examined. Each bp has four values, therefore the probability of a mismatch when two random bps are compared is ¾. Comparing two random sections of equal length from DNA sequences leads, on average, to 75% mismatching bps. However, when high similarity fragments are compared, the running Hamming distance may drop significantly below the 75% average, thus indicating a potential mapping location.

RASSA is a resistive memory based massively parallel processing-in-memory accelerator. It allows storing (typically, a data element per memory row) and in-situ processing of large datasets. RASSA enables comparing a key pattern with the entire dataset in parallel. Every number of mismatches (of the key pattern vs. each data element that is in each memory row) causes a specific voltage drop, allowing quantifying the number of mismatching locations (called a *mismatch score*). The mismatch score is compared with a predefined threshold value to detect the locations which have the desired degree of similarity with the compared pattern, indicating a viable mapping location. The following sections describe RASSA functionality, encoding of DNA bp, RASSA system architecture and hardware evaluation.

## DNA Base Pair Encoding and Mismatch Evaluation

Figure 1b presents the RASSA bitcell, containing two transistors and one memristor (2T1R). Each memristor serves as a single bit storage element and a single bit comparator, enabled by the selector transistor.

A compare operation consists of two phases, the precharge and the evaluation. During precharge, the Match Line is precharged to a certain voltage level. At the same time, the evaluation transistor in each bitcell is on, to discharge the evaluation point (created by the diffusion capacitances of the selector and the evaluation transistors).

During the evaluation phase, if the selector transistor is on, a low memristor resistance ($R_{ON}$) allows charge to pass from the match line to the evaluation point. The charge distribution causes the match line voltage to drop. Sensing the voltage of the match line compared with a reference voltage (of zero mismatch) allows quantifying the number of mismatches, producing the mismatch score.

## Encoding

RASSA reserves four bitcells to store a DNA bp. There are four nucleotide bases, A, C, G and T in each DNA bp, encoded using one-hot encoding as '1000', '0100', '0010' and '0001', respectively. While it is possible to encode four nucleotide bases using two bits (for example '00', '01', '10', and '11' for A, C, G and T, respectively), such encoding would result in different number of mismatching bits depending on a specific pair (for example, two mismatching bits in the case of A-T or C-G mismatch, or one mismatching bit in the case of A-C or A-G mismatch), leading to ambiguous results. Since a mismatch is signaled by reduced match line voltage (caused by charge redistribution), a match should block charge flow. One-hot encoding assures that at most one mismatch may happen in each group of four bitcells. For instance, in 60 bitcells, at most 15 mismatches may be observed. Therefore, in this work, a memristor in high resistive state ($R_{OFF}$) is considered logic '1' while $R_{ON}$ is considered logic '0'.

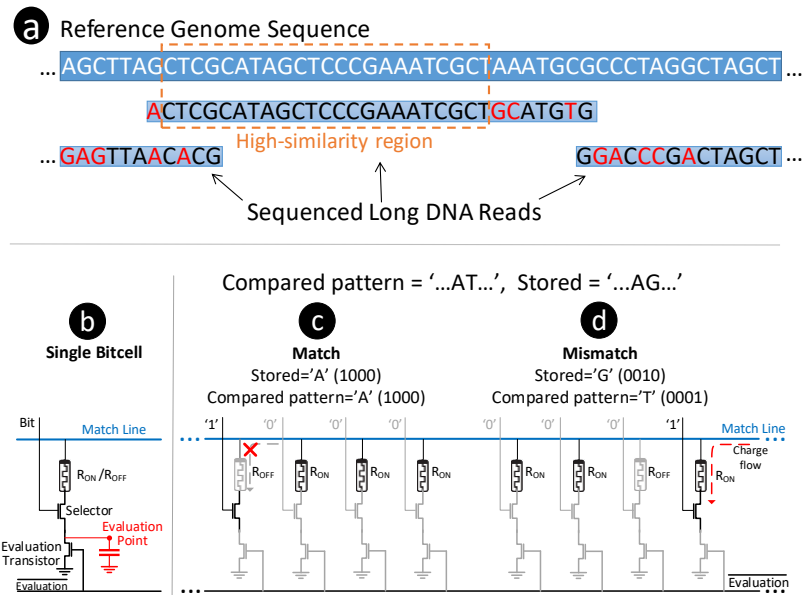

Figure 1. (a) Mapping of long DNA reads onto existing reference sequence. Red colored bps represent mismatching bps between the reference and reads. (b) Single RASSA bitcell. (c), (d) example of two DNA bps comparison. One bp matches the compared pattern, preventing match line charge loss (c). The next bp mismatches, causing match line voltage reduction (d).

## Mismatch Evaluation

During a compare operation, the compared (key) pattern is applied to the gates of the selector transistors of all bitcells. If certain groups of bitcells need to be ignored (masked-out) during comparison, zero is applied to the gates of the selector transistors of such bitcells. Figure 1c shows a stored 'A' nucleotide symbol and a compare pattern of 'A'. The comparison results in a match, so there is no charge redistribution path (through an $R_{OFF}$ memristor). Figure 1d shows a mismatch, where the stored pattern is 'G' and the key pattern is 'T'. The mismatch results in charge redistribution through an $R_{ON}$ memristor, causing a match line voltage drop.

Figure 2a shows all possible match line voltage levels during the evaluation phase for mismatch scores of 0 through 15. The match line is sensed by an analog-to-digital converter (ADC, "System Architecture"

Section). The timing of such sensing, in addition to the per-cell transistor capacitance variations, may lead to inaccuracies in the mismatch score. For example, for a match line shared by up to 60 bitcells, the mismatch score error could be $\pm 1$. If the number of bitcells sharing the match line is more than 120, the mismatch score error could reach $\pm 3$.

## System Architecture

The main component of RASSA is the 2T1R array, divided into Word Rows (Figure 2c), further divided into Sub-Words (Figure 2b). All Word Rows are connected in parallel to the Key Pattern register (Figure 2d). The ADC is the largest and most energy consuming component of a Sub-Word. Therefore, in order to use only 4-bit ADC, supporting mismatch scores of 0 through 15, the Sub-Word is limited to 60 bitcells. There are 16 Sub-Words within a Word Row, amounting to 960 bitcells per word, designed for storing and comparing up to 240 DNA bps per cycle. In each compare operation, a compare pattern is applied to all active bitcell bit lines.

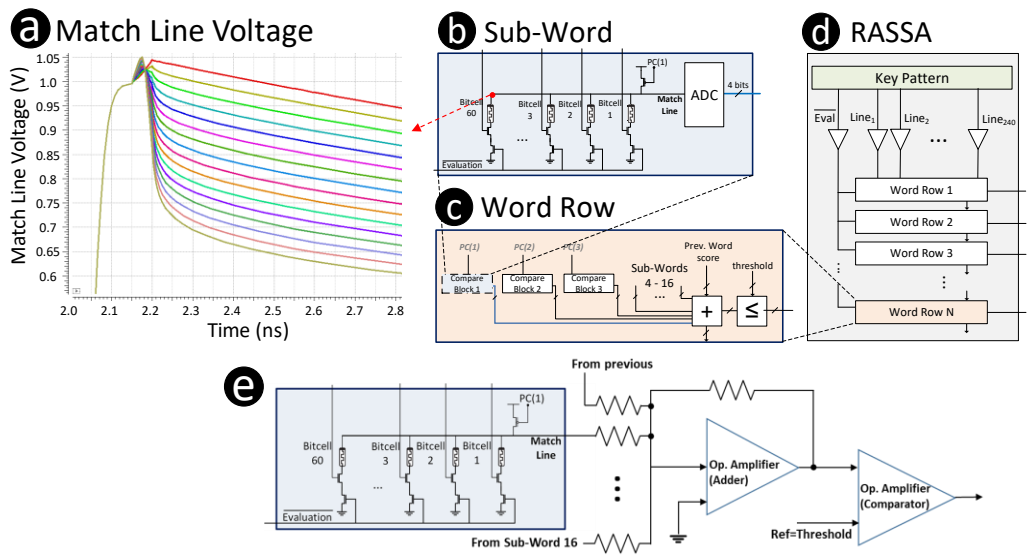

Figure 2. (a) Match line voltage levels for each mismatch score between zero (top curve) and 15 (bottom curve) mismatches. Every voltage level at the sampling point is converted by the ADC. (b-d) Bottom-up block diagram of RASSA. (b) Single Sub-Word, composed of 60 bitcells in NOR-like structure. (c) Single Word Row containing 16 Sub-Words, capable of holding 240 DNA bps. (d) Complete RASSA diagram containing N Word Rows (N=$2^{17}$). (e) Accumulating mismatch scores in analog domain (Conclusions and Future Research Directions Section)

The match line voltage of each Sub-Word is sampled by the ADC and converted into a 4-bit mismatch score (right side of b). The ADC reference voltage and voltage level differences are set according to the match line values for each mismatch score, as demonstrated in Figure 2a. The 16 Sub-Word ADC outputs are summed up to produce the mismatch score for the entire Word Row (Figure 2b). All such scores are then compared with a threshold value, in parallel, to indicate the Word Rows (corresponding to sequence locations) with the desired degree of similarity.

## Timing, Power and Area Breakdown

A Sub-Word circuit is designed, placed and routed using the 28nm CMOS High-k Metal Gate library from Global Foundries for transistor sizing, timing and power analysis. We perform Spectre simulations for the FF and SS corners at $70^0C$ and nominal voltage. Timing analysis show an operational frequency of 1GHz is possible. For a single Sub-Word, the precharge energy is 1.6fJ, while the evaluation energy (ADC and control line switching) is 98.4fJ. For a single Word, containing 16 Sub-Words, adders and threshold comparator, a single compare cycle energy is 1791fJ.

We have manually laid-out a RASSA bitcell. The total Word Row area in 28nm technology, including the bitcells, ADC, adders and comparator is $1598\mu m^2$. Bitcell transistors occupy 4%, adders and threshold comparator occupy 28% and the ADC occupies 68% of the Word Row area, respectively. This allows placing of 131k ($2^{17}$) 960-bit (240-bp) Word Rows, storing 31.5Mbps, on a single $209 mm^2$ die. Its worst case power consumption at 1GHz is 235W. Table 1 summarizes the RASSA system parameters.

Table 1. RASSA System Parameters for 28nm node process

| Parameter | Value |
| --- | --- |
| DNA bps per row (bits) | 240 (960) |
| Words per IC | $131k$ ($2^{17}$) |
| Memory size (DNA bps) | $31.5M$ |
| Frequency | $1GHz$ |
| Single IC Power | 235W |
| Single IC Area | $209 mm^2$ |

Loading the reference sequence to RASSA is performed on each Word Row separately and requires two cycles per Word Row, one cycle to write all logic '0's, and another cycle to write all logic '1's. Given a reference sequence of length $L$, the number of cycles required for its loading equals $2 \cdot \left\lceil \frac{L}{240} \right\rceil$.

## DNA READ PRE-ALIGNMENT FILTERING WITH RASSA

A single compare operation in RASSA finds the mismatch score between the key pattern and the contents of each Word Row. The reference sequence is stored in RASSA (contiguously, 240-bp fragment per Word Row). A fixed-size chunk (e.g., 200 bps) of the read is fed in as a key pattern. The mismatch score approximates the correlation between the read chunk and the reference sequence. A long read contains multiple chunks, therefore the compare operations are performed multiple times, in all possible positions of a read chunk vis-à-vis a Word Row reference fragment, sometimes involving two neighboring Word Rows.

The number of Word Rows in RASSA defines the number of overlap positions examined simultaneously. In a single cycle, $\lceil n/240 \rceil$ (where $n$ is the reference sequence length) distinct positions on the reference sequence are examined simultaneously. To cover all possible positions, the read chunk is shifted by one bp, and compare is repeated 240 times (resembling the concept of correlation).

Figure 3a-d illustrates the comparison of a read chunk against a reference sequence in RASSA for several cases. In these examples, a chunk length of 200 bp is used (Figure 3a). A multi-cycle compare operation matches a 200bp chunk against all its possible locations vis-à-vis the reference sequence. In the first compare cycle (Figure 3b), the first chunk of the read is compared with all first 200 bps of each RASSA Word Row.

Following the completion of 41(=240-200+1) cycles, the 200bp chunk is compared against reference data residing in two Word Rows. Such two-Word Row compare requires two cycles. The even cycle mismatch score (Figure 3c) is added to the score of the following odd cycle (of the Word Row below, Figure 3d), and compared with the threshold (Figure 3d, right). Before every even cycle, the compare pattern is shifted by one bp to the right, shortening the even cycle compare pattern and extending the pattern in the odd cycle by one bp (Figure 3c,d, right). After 439 (=41+199×2) cycles, a 200 bp chunk has been compared against all reference sequence positions. The compare operation repeats for the rest of the 200bp read chunks.

Figure 3e presents the concept of edit detection in RASSA. For simplicity, the chunk length in this example is 30bp. Three types of edits are shown on the left of the figure. On the right, the mismatch score is presented for all relative shifts of the chunk versus the reference.

The average mismatch score in any mismatching location is 75% (the probability of an individual bp mismatch is 0.75). In the exact match case (top of the figure), the mismatch score is 0.

A substitution results in a very low mismatch score easily detectable by RASSA. The second row of Figure 3e shows two substitution errors, leading to the mismatch score of 2/30=6.7%.

The third row of Figure 3e shows an insertion error. The longest matching section is 18 bp to the right of the insertion, which leads to a mismatch score of (30-18)/30=40%. With the appropriate threshold[*], such scenario is detectable by RASSA and identified as a potential mapping.

Last, in the fourth row of Figure 3e, a deletion error is presented. Deletions are handled similarly to insertions. In this example, the longest matching section is also 18 bp. The mismatch score is (30-18)/30=40% as well, and is also detectable by RASSA as a potential mapping.

While a substitution results in a much lower mismatch score, RASSA is capable of detecting indels just as confidently, by setting the threshold accordingly[*].

Figure 3f,g illustrates the mismatch score for the sliding window search of RASSA, in presence of multiple errors. Figure 3f presents an example of a reference sequence and a read chunk containing a high-similarity region. All possible edit types (substitution, insertion and deletion) exist in the chunk. Figure 3g illustrates the mismatch score as a function of the compare cycle (the relative read chunk position). During most cycles, the mismatch score is above threshold. When the chunk is in its valid mapping location ("min mismatch position" at cycle 9 in Figure 3f), the mismatch score is significantly lower than the random average 75% level. Setting the custom threshold, for instance, at 50% allows efficient overlap of read chunks with a number of edits (substitutions as well as indels).

## Translating the Output of RASSA to Mapping Locations

RASSA compares every chunk of every read against the entire reference sequence. The probability of a false positive match is extremely low[*]. Therefore, we assume that every compare that results in a mismatch score below the predefined threshold indicates is valid mapping location for the entire read. The output of RASSA is a bit vector, one bit per Word Row. The index of the Word Row, together with the iteration number and relative position of the chunk within the read, provide an exact coordinate of a potential mapping location. In most examined cases, a single read has a single mapping location indicated by a single compare from a single chunk.

In some other cases, a single or multiple chunks produce multiple potential mapping locations. In such cases, the distance between consecutive locations is examined, starting from the lowest coordinate. If the distance between two consecutive locations is smaller than the read length, the location associated with the higher coordinate is discarded. Otherwise, both locations are kept for further processing. With this selection heuristic, nearby potential mapping locations from a single or multiple chunks are combined, while distant locations are treated as separate mapping locations.

The mapping locations identified by RASSA can be further verified by alignment (e.g., Smith-Waterman algorithm[4]), and used by assembly or error correction programs[8]. Unmapped reads can either be discarded (in case of a high sequencing coverage) or be mapped with a seed-and-extend mapper, and then verified by an alignment algorithm.

---

[*] In legitimate mismatching locations, the mismatch score is distributed binomially. The threshold is set such that the probability of the mismatch score to fall below it is sufficiently low. For example, P(mismatch score<50%)=0.0008 for 30bp chunk. For 200bp chunk, similar probability is reached at the threshold of 65%.

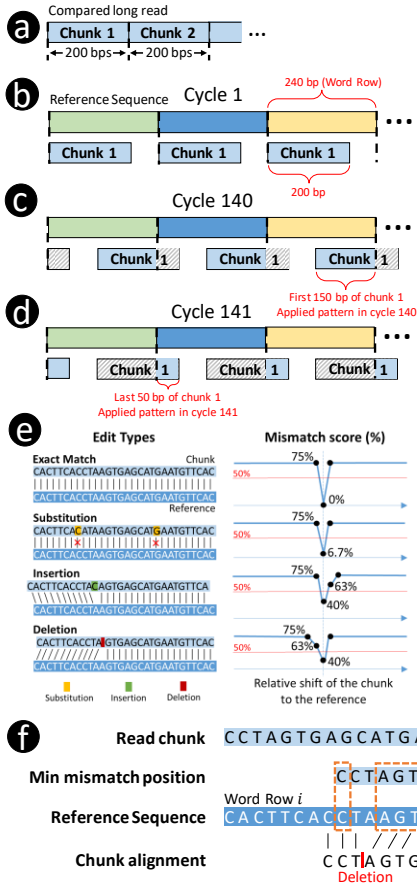
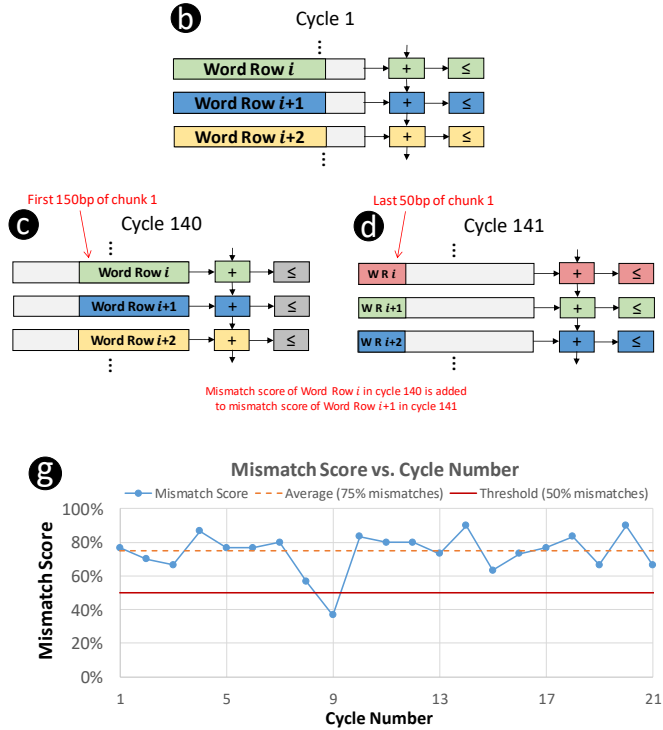

Figure 3. Illustration of a single long read chunk examination in RASSA. (a) A long read is divided to chunks, each 200 bp long. (b, left and right) First chunk is compared against the reference sequence in multiple locations (simultaneously); (c),(d) First chunk overlaps with reference sequence bps from two Word Rows; (c, left) The first part of the chunk compared with the last bps of the Word Row. (c, right) All Sub-Word mismatch scores are summed up and stored (compare to threshold does not take place); (d, left) Second part of the chunk is compared with the first bps of the next Word row; (d, right) All Sub-Word mismatch scores, including the previous cycle result from the above Word Row, are summed up and compared with a threshold. Following this step, the chunk is shifted right by one position (relative to the reference) and steps (c) and (d) are repeated. (e) Edit types and the mismatch score found by RASSA. (f) Example of a 30-bp read chunk containing insertion, deletion and substitution errors is compared against the reference sequence divided into 50-bp Word Rows; (g) Mismatch score vs. cycle number for first 21 cycles of comparison of the example in (f). Minimal mismatch score is below the threshold and achieved in the 9[th] cycles. The threshold is determined empirically per dataset.

# EVALUATION

We compare RASSA with two existing solutions, minimap2[7], a state-of-art read mapping tool, and GateKeeper[9], a state-of-art short read pre-alignment hardware accelerator.

## Comparison with Minimap2

Our evaluation focuses on accuracy and speedup. Accuracy is measured by two criteria: (1) *sensitivity*: correctly mapped reads, (2) *false positives*: percentage of incorrect mappings out of all mappings by RASSA.

### Methodology

Minimap2 is run with the parameters '-x map-pb' and '-x map-ont', invoking its execution for overlap detection without the alignment step. With such parameters, minimap2 functions as a pre-alignment filter. The parameters also invoke appropriate heuristic for PacBio and ONT reads, in addition to enabling multi-threading and SIMD extensions. To evaluate RASSA's pre-alignment filtering accuracy, we use minimap2 as a *golden reference*. Speedup is calculated as the ratio of minimap2 execution time, without indexing, to RASSA execution time. The accuracy and speedup of RASSA were obtained using an in-house simulator. We assume that the reference sequence has already been loaded into RASSA prior to execution.

To find the number of incorrect output locations which might increase total alignment time we have contrasted pre-alignment followed by alignment with alignment without pre-alignment. We have used part of the E.coli PacBio dataset, consisting of about 1000 reads. Total pre-alignment by RASSA took 20msec, and the following alignment needed to be applied to only 70kbp subset of the reference, taking minimap2 1490msec. In contrast, the same alignment applied without the pre-alignment stage took 3000msec, about twice the time. Therefore, we decided that reads with more than two output locations by RASSA will be discarded and treated as incorrectly mapped.

### Datasets

We use five publicly available datasets, three from PacBio and two from ONT, taken from two organisms: E.coli K-12 NG1655 and Saccharomyces cerevisiae W303 (yeast). Both reference sequences are available at the NCBI (https://www.ncbi.nlm.nih.gov/). PacBio datasets were taken from https://github.com/Pacific-Biosciences/DevNet/wiki/Datasets. Error rates[13], including the share of insertions, deletions and mismatches (I,D,M) are also presented.

**E.coli**
- PacBio: 100k reads from one SMRT cell, 5245 bps on average.
  Error rate: 14.2% (I:41.7%, D:21.2%, M:37.1%)
- PacBio CCS: 260k high-quality CCS reads from 16 SMRT cells, 940 bps on average.
  Error rate: 1% (I:5%, D:19.5%, M:75.5%)
- ONT (from http://lab.loman.net/2016/07/30/nanopore-r9-data-release/): 165k R9 1D reads, 9009 bps on average. Error rate: 20.2% (I:14.5%, D:37.2%, M:48.3%)

**Yeast**
- PacBio: 100k reads from one SMRT cell, 6294 bps on average.
  Error rate: 14% (I:5%, D:19.5%, M:75.5%)
- ONT (ERR789757 from NCBI): 30k R7.3 2D MinION reads, 11337 bps on average.
  Error rate: 13.4% (I:23.3%, D:35.7%, M:41%)

Table 2 presents the accuracy results for all five datasets above. Chunk sizes of 200 and 100 bps and corresponding thresholds were determined empricially, trading off accuracy and performance. Small changes of threshold induce only marginal changes in accuracy. For most datasets, 55% threshold was used on 200bp chunks and 45% for 100pb chunks; for the Yeast PacBio case, we used 45% and 40%, respectively.

Table 2. Sensitivity, fraction of exact mappings and speedup of RASSA compared to minimap2

| Datasets | Large Chunk (200 bps) | | | Small Chunk (100 bps) | | |
|---|---|---|---|---|---|---|
| | Sensitivity | False Positives | Speedup | Sensitivity | False Positives | Speedup |
| E.coli PacBio | 79.3% | 13.4% | 25x | 83.2% | 13.6% | 16x |
| E.coli PacBio CCS | 96.3% | 8.9% | 43x | 96.2% | 6.9% | 24x |
| E.coli ONT | 88.8% | 10.5% | 48x | 87.6% | 12.4% | 31x |
| Yeast PacBio | 69.8% | 8.7% | 77x | 72% | 11.8% | 51x |
| Yeast ONT* | 85.9% | 34.9% | 31x | 85.1% | 39.2% | 49x |

* minimap2 mapped only about 20% of all reads, with 50% of mappings with lower quality score than 60 (indicates a high-confidence mapping).

## Speedup

We compare RASSA execution time with that of minimap2, executed on a server with 16-core 2GHz Intel Xeon E5-2650 CPU and 64GB of RAM. Table 2 shows that RASSA achieves 16-77× speedup over minimap2[†]. We note that the yeast dataset has fewer reads than E.coli, but a longer reference sequence (11.7Mbp vs. 4.6Mbp), which might cause the longer execution time on minimap2. In contrast, RASSA is insensitive to the reference sequence length and its execution time is determined by the length of a read chunk.

RASSA produces output (on average, one mapping per read) at rate of 50,000—500,000 reads/sec, enabling multiple simultaneously executing instances of a typical alignment algorithm.

## Throughput Comparison with GateKeeper

We compare RASSA throughput (the number of examined mapping locations per second) with that of GateKeeper[9]. GateKeeper was implemented in a Virtex-7 FPGA using Xilinx VC709 board running at 250MHz.

GateKeeper is designed to compare short reads with a reference sequence. Table 3 shows the throughput in Billions of Examined Mapping Locations per second, BEML/s, of RASSA and GateKeeper on two short read datasets used in [9]: (1) 100bp reads (2) 300bp reads. RASSA frequency is adjusted to that of GateKeeper. In addition, we show the average RASSA throughput for the 200bp reads, equivalent to the chunks lengths used in Table 2.

RASSA outperforms GateKeeper by more than 2 order of magnitude. When applied to short read mapping pre-alignment, RASSA covers a read with one chunk. Consequently, RASSA takes 1-3 cycles (for the read lengths used in Table 3) to find the mismatch score in all Word Rows in parallel.

Table 3. RASSA and GateKeeper[9] throughput (billions of examined mapping locations per second, BEML/s)

| Read Lengths | GateKeeper | RASSA @250MHz |
|---|---|---|
| 100bp | 1.7 BEML/s | 226.8 BEML/s |
| 200bp | - | 175.2 BEML/s |
| 300bp | 0.2 BEML/s | 142.8 BEML/s |

[†] The overlap detection stage without alignment of minimap2

GakeKeeper, on the other hand, is reported to process up to 140 (20) mapping locations of 100bp (300bp) reads in parallel, affected by the edit distance threshold.

# CONCLUSIONS AND FUTURE RESEARCH DIRECTIONS

This paper presents RASSA, an in-memory processing parallel architecture of a Resistive Approximate Similarity Search Accelerator. We apply RASSA to the long read DNA mapping problem. The length of reads, coupled with a low read quality, poses a challenge for existing mappers, optimized for high quality short reads. The read mapping process is data and compute intensive, making it a target for acceleration. RASSA addresses the challenge by breaking long reads into short chunks and by applying full correlation. By allowing faster mapping on large datasets, we potentially make a step towards real time pathogen or genome sequence completion.

We compared RASSA accuracy and execution time with that of minimap2, a state-of-the-art mapping solution, on five long read datasets taken from two organisms. Our evaluation shows that RASSA can outperform minimap2 by 16-77×. In addition, we compared RASSA's throughput, measured in examined mapping locations per second, with that of GateKeeper, a state-of-art short read pre-alignment hardware accelerator. We find that RASSA can outperform GateKeeper by more than 2 orders of magnitude.

This work can be extended in several ways. First, RASSA can be applied to read-to-read overlap finding, which requires finding overlaps between pairs of reads. Read-to-read overlap finding is an important first step in *de novo* genome assembly[6] (constructing the host DNA sequence without a reference sequence), a problem more computationally challenging than read mapping. Second, a detailed design space exploration needs to be performed. For example, RASSA can be further optimized in terms of hardware cost: higher density can be achieved by sharing analog to digital converters among multiple Sub-Words and by applying analog computations, as presented in Figure 2e. RASSA mapping and resistive CAM alignment[5] may be combined into a single high performance in-memory mapper/aligner. Last, thanks to its use of short chunks, RASSA can be effectively applied to short reads.

## REFERENCES

1. Jameson, J.L. and Longo, D.L. "Precision medicine—personalized, problematic, and promising." Obstetrical & gynecological survey, vol. 70, no. 10, pp. 612-614, 2015.
2. Quick, J., Loman, N.J., Duraffour, S., et al, "Real-time, portable genome sequencing for Ebola surveillance." Nature, 530(7589), pp. 228-232.
3. Rhoads, Anthony, and Kin Fai Auc. "PacBio sequencing and its applications." Genomics, proteomics & bioinformatics 13.5, pp. 278-289, 2015.
4. Laver, T., Harrison, J., O'neill, P. A., Moore, K., Farbos, A., Paszkiewicz, K., & Studholme, D. J. "Assessing the performance of the oxford nanopore technologies minion." Biomolecular detection and quantification, vol. 3, pp. 1-8, 2015.
5. Kaplan, R., Yavits, L., Ginosar, R. and Weiser, U. "A Resistive CAM Processing-in-Storage Architecture for DNA Sequence Alignment." IEEE Micro, vol. 37, no. 4, pp. 20-28, 2017.
6. Li, Heng. "Minimap and miniasm: fast mapping and de novo assembly for noisy long sequences." Bioinformatics 32.14, pp. 2103-2110, 2016.
7. Li, H., Minimap2: pairwise alignment for nucleotide sequences. Bioinformatics, vol. 1, p. 7, 2018.
8. Berlin, K., Koren, S., Chin, C. S., Drake, J. P., Landolin, J. M., & Phillippy, A. M. "Assembling large genomes with single-molecule sequencing and locality-sensitive hashing." Nature biotechnology, 33(6), p. 623, 2015.
9. Alser, M., Hassan, H., Xin, H., Ergin, O., Mutlu, O. and Alkan, C. "GateKeeper: a new hardware architecture for accelerating pre-alignment in DNA short read mapping." Bioinformatics, vol. 33, no. 21, pp. 3355-3363, 2017.
10. Kim, Jeremie S., et al. "GRIM-Filter: Fast seed location filtering in DNA read mapping using processing-in-memory technologies." *BMC genomics* 19.2 (2018): 89.

11. Imani, M., Rahimi, A., Kong, D., Rosing, T., & Rabaey, J. M. "Exploring hyperdimensional associative memory". In IEEE International Symposium on High Performance Computer Architecture (HPCA), pp. 445-456, 2017.
12. Akinaga, Hiroyuki, and Hisashi Shima. "Resistive random access memory (ReRAM) based on metal oxides." Proceedings of the IEEE 98.12, pp. 2237-2251, 2010.
13. Weirather, Jason L., et al. "Comprehensive comparison of Pacific Biosciences and Oxford Nanopore Technologies and their applications to transcriptome analysis." *F1000Research, vol* 6, 2017.

## ABOUT THE AUTHORS

**Roman Kaplan** received his BSC and MSc from the faculty of Electrical Engineering, Technion, Israel in 2009 and 2015, respectively. Between 2009-2014 he worked as a software engineer. He is a PhD candidate in the faculty of Electrical Engineering, Technion, under the supervision of Prof. Ran Ginosar. Kaplan's research interests are parallel computer architectures, in-data accelerators for machine learning and bioinformatics.

**Leonid Yavits** received his MSc and PhD in Electrical Engineering from the Technion. After graduating, he co-founded VisionTech where he co-designed a single chip MPEG2 codec. VisionTech was acquired by Broadcom in 2001. Leonid is currently a postdoc fellow in Electrical Engineering in the Technion. He co-authored a number of patents and research papers on SoC and ASIC. His research interests include non-von Neumann computer architectures, accelerators and processing in memory.

**Ran Ginosar** (M'78) received his BSc from the Technion—Israel Institute of Technology in 1978 (summa cum laude) and his PhD from Princeton University, USA, in 1982, both in Electrical and Computer Engineering. He joined the Technion faculty in 1983 and was a visiting Associate Professor with the University of Utah in 1989-1990, and a visiting faculty with Intel Research Labs in 1997-1999. He is a Professor at the Department of Electrical Engineering and serves as Head of the VLSI Systems Research Center at the Technion. His research interests include VLSI architecture, manycore computers, asynchronous logic and synchronization, networks on chip and biologic implant chips. He has co-founded several companies in various areas of VLSI systems.